%% file: main.tex
\documentclass[letterpaper, 10pt, conference]{ieeeconf}
\IEEEoverridecommandlockouts
\usepackage{graphics} %
\usepackage{epsfig} %
\usepackage{times} %
\usepackage{amsmath} %
\usepackage{amssymb}  %
\usepackage{epstopdf}
\usepackage{cite}
\usepackage{mathtools}
\usepackage[caption=false, font=footnotesize]{subfig}

\usepackage{color}
\usepackage{booktabs}
\usepackage{fancyhdr}
\usepackage[short]{optidef}
\usepackage{lipsum}
\usepackage{algpseudocode}
\usepackage{algorithm}

\input{notation.tex}

\usepackage[dvipsnames]{xcolor}
\usepackage[deletedmarkup=sout,authormarkup=superscript]{changes}
\definechangesauthor[name={Samuel}, color=NavyBlue]{SB}
\definechangesauthor[name={Dominic}, color=RedViolet]{DLM}
\definechangesauthor[name={Alisa}, color=Orange]{AR}
\definechangesauthor[name={ToDo}, color=Green]{TODO}
\definechangesauthor[name={Efe}, color=OliveGreen]{EB}
\newcommand{\sam}[1] {\added[id=SB]{[#1]}}
\newcommand{\eb}[1] {\added[id=EB]{[#1]}}

\title{\LARGE \bf
Sequential Quadratic Programming-based\\ Iterative Learning Control for Nonlinear Systems
}

\author{Samuel Balula$^{1,3}$, Efe C. Balta$^{1}$, Dominic Liao-McPherson$^{2}$, Alisa Rupenyan$^{1,3}$, and John Lygeros$^{1}$%
\thanks{This project has been funded by the Swiss Innovation Agency (Innosuisse, Grant Number 46716) and the Swiss National Science Foundation through NCCR Automation, a National Centre of Competence in Research (Grant Number 180545).}
\thanks{$^{1}$ S. Balula, E. C. Balta, A.  Rupenyan, and J. Lygeros are with the ETH Z\"{u}rich Automatic Control Laboratory, Physikstrasse 3, 8092 Z\"{u}rich, Switzerland. E-mail: \texttt{\{sbalula, ebalta, ralisa, jlygeros\}@ethz.ch}.}
\thanks{$^{2}$ D. Liao-McPherson is with the University of British Columbia, 2054-6250 Applied Science Ln, Vancouver, BC V6T 1Z4, Canada. Email: \texttt{dliaomcp@mech.ubc.ca}.}
\thanks{$^{3}$ S. Balula and A. Rupenyan are also with Inspire AG.}
}
\begin{document}

\maketitle

\begin{abstract}
	\input{src/0-abstract.tex}
\end{abstract}

\markboth{IEEE Conference on Control Technology and Applications}%
{}

\section{Introduction} 
\input{src/1-introduction.tex}

\section{Problem Setting}
\label{sec:problem}
\input{src/2-problem.tex}

\section{Optimization Based ILC for non-linear systems}
\label{sec:ob-ilc}
\input{src/3-method.tex}

\section{Case study with a 2D precision motion system}
\label{sec:case-study}
\input{src/4-case-study.tex}

\input{src/5-results.tex}

\section{Conclusion}
\label{sec:conclusion}
\input{src/6-conclusion.tex}

\bibliography{database}

\end{document}

%% file: notation.tex
\newcommand{\tp}{^\mathsf{T}}

\newcommand{\Reals}{\mathbb{R}}

\newcommand{\Target}{\Xi}
\newcommand{\Lagrangian}{\mathcal{L}}

\newcommand{\dz}{\begin{bmatrix} \Delta u \\ \Delta p \end{bmatrix}}

%% file: src/0-abstract.tex
Learning-based control methods for industrial processes leverage the repetitive nature of the underlying process to learn optimal inputs for the system. While many works focus on linear systems, real-world problems involve nonlinear dynamics. In this work, we propose an algorithm for the nonlinear iterative learning control problem based on sequential quadratic programming, a well-studied method for nonconvex optimization. We repeatedly solve quadratic subproblems built using approximate nonlinear models and process measurements, to find an optimal input for the original system. We demonstrate our method in a trajectory optimization problem for a precision motion system. We present  simulations to illustrate the performance of the proposed method for linear and nonlinear dynamics models.

%% file: src/1-introduction.tex
Iterative learning control (ILC) is used in repetitive tasks to improve performance over iterations by learning from previous trials. In ILC, the control input is updated between iterations using the measured error, which is shown to ensure monotonic convergence to an approximate fixed point of the original problem under various assumptions~\cite{liao2022robustness,barton2010norm,son2015robust,tayebi2007unified}.  

An important challenge with ILC is to ensure convergence and constraint satisfaction, which is especially difficult when the underlying system is nonlinear. 
Optimization-based ILC (OB-ILC) methods have been proposed in the literature to systematically study iteration-wise error dynamics and constraint satisfaction.
Robust optimization-based methods~\cite{adlakha2020optimization,son2015robust}, interior point-based OB-ILC~\cite{mishra2010optimization}, and norm-optimal ILC methods~\cite{amann1996iterative,barton2010norm,balta2021learning,gunnarsson2001design} are some of the common approaches in the literature for linear systems.
While some of the works consider model mismatch and process constraints jointly, many of the existing works do not provide robust constraint satisfaction, and convergence results in the presence of measurement noise. 
Recently, OB-ILC has been extended to handle process constraints in linear processes while accounting for noise and model mismatch during all iterations~\cite{liao2022robustness}.

This work aims to extend existing OB-ILC methods to nonlinear system dynamics.
Specifically, our goal is to leverage approximate process models to pose an optimization problem that we iteratively solve using the underlying nonlinear system while ensuring constraint satisfaction.

A survey of the ILC method for nonlinear dynamics is given in~\cite{xu2011survey}.
In~\cite{tayebi2007unified} robust convergence for a class of nonlinear systems is given, while a neural network-based nonlinear ILC method is presented in~\cite{yu2021neural}.%
Linearization-based OB-ILC methods for nonlinear systems are studied in~\cite{schollig2009optimization,lu2017nonlinear}. 
Variants of Newton-based methods are used for nonlinear ILC problems~\cite{avrachenkov1998iterative,lin2006newton,volckaert2009model}.
In~\cite {baumgartner2020zero} a zeroth-order ILC for nonlinear processes is proposed. It requires solving a nonlinear program after each iteration and difficult-to-verify properties with approximate sensitivities. 
In this paper, we consider a similar setting but propose a novel nonlinear OB-ILC method based on the well-known sequential quadratic programming (SQP) method for nonconvex optimization~\cite{boggs1995sequential}. 
Specifically, we consider model mismatch and constraints to form approximate subproblems, which are solved by using measurements from the nonlinear process. 
The main contribution of this paper is a nonlinear OB-ILC scheme based on the SQP framework, that requires solving convex quadratic subproblems after each trial and can handle constraints and approximate models.

ILC is used extensively in motion tracking problems and has been shown to improve the performance of gantry systems~\cite{chen2021iterative}, wafer stages~\cite{mishra2010optimization}, precision motion systems~\cite{balta2021learning,barton2010norm} and various related applications~\cite{bolder2014rational,bristow2006high}.
Similarly, we illustrate our proposed OB-ILC method for nonlinear dynamics on a precision motion tracking problem. 
We present a detailed case study using a high-fidelity simulator of a precision motion system, and we compare the achieved tracking accuracy by using models with different fidelity (linear and neural network-based).

The rest of the paper is structured as follows. Section~\ref{sec:problem} presents the problem setting and the control approach. Section~\ref{sec:ob-ilc} presents the optimization problem and the proposed OB-ILC approach. Section~\ref{sec:case-study} presents a detailed case study in precision motion control and Section~\ref{sec:conclusion} provides closing remarks with potential future directions.

We denote the Jacobian by $\nabla$ and the Jacobian along a certain direction $d$ by $\nabla_d$.
Similarly $\nabla^2$ and $\nabla^2_d$ are the Hessian and the Hessian along the direction $d$ respectively. $\partial^n$ denotes the discrete derivative of order $n$, defined by $\partial^n x =(\partial^{n-1} x(i+1) - \partial^{n-1} x(i))/{\Delta t}$, $\partial^0 x = x$, where $\Delta t$ is the discrete time interval.

\if01
\sam{Notation:\\
$u_k$: Input \\
$y_k$: Measurement\\
$p_k$: Local estimate of $f(u_k)$ \\
$z_k = \begin{bmatrix} u_k \\ p_k \end{bmatrix}$: Collection of variables\\
$x_k = \begin{bmatrix} z_k \\ \lambda \\ \sigma \end{bmatrix} = \begin{bmatrix} \begin{bmatrix} u_k \\ p_k \end{bmatrix} \\ \lambda \\ \sigma \end{bmatrix} $: ILC policy internal state\\
$\eta_k = \frac{\eta_0}{\sqrt{k}}$: damping coeficient "SQP / Newton style".\\
$\mathcal{T}$: ILC policy.
}
\fi

%% file: src/2-problem.tex
We consider a noisy nonlinear repetitive process of the form
\begin{equation}
	y = f(u) + w,
 \label{eq:repetitive-process}
\end{equation}
where $u\in\Reals^n$ is the input,
$y\in\Reals^m$ is the output,
$f:\Reals^n\rightarrow\Reals^m$ is the system response (input/output map), and
$w\in\Reals^m$ is a non-repeating disturbance, assumed to be zero-mean. %
We focus on the response of a dynamical system over a finite interval where $u$ and $y$ define input/output trajectories of the underlying nonlinear system.
Therefore, we have $u=\left(u(1), u(2), \ldots,u(N)\right)$ for an input trajectory of $N$ time steps, and similarly for $y$.
For example, the input $u$ could be a trajectory of actuator commands or a reference trajectory tracked by a low-level feedback controller, and the output $y$ is the actual trajectory traced by the system. In Section \ref{sec:case-study} we show a case study with this configuration.
The input and output must satisfy the constraints $u \in \mathcal{U} \subset \Reals^n$, $y \in \mathcal{Y} \subset \Reals^m$. 
For example, in a motion tracking problem, the sets $\mathcal{U}$ and $\mathcal{Y}$ may encode limits on actuation velocity, and acceleration. 

Our control objective is to choose the input $u$ such that the output $y$ tracks a target trajectory $\Target$ as closely as possible. The ILC approach designs a learning policy $\pi = (x, \mathcal{T}, q)$ of the form 
\begin{subequations}
	\begin{align}
		x_{k+1} &=\mathcal{T}(x_k,y_k),\\
		u_k   &= q(x_k) .
	\end{align}
\end{subequations}
where $x$ is the internal state of the policy,
$\mathcal{T}$ is the update function,
and $q$ is an output function that recovers the control input from $x$.
The goal is to design $x$, $\mathcal{T}$, and $q$ such that the (iteration domain) closed-loop system
\begin{subequations}
\label{eq:ideal_policy}
	\begin{align}
		u_k &= q(x_k), \\
		y_k &= f(u_k) + w_k, \\
		x_{k+1} &= \mathcal{T}(x_k, y_k),
	\end{align}
\end{subequations}
converges to some $y_k$ close to $\Target$, with  $u \in \mathcal{U}$ and $y \in \mathcal{Y}$. Due to the dynamics and constraints $y_k$ will, in general, not reach $\Target$ exactly.
Subscript $k$ indicates the iteration index of the ILC throughout the rest of the chapter.

\begin{figure}[t]
	\begin{center}
    \vspace{0.2cm}
		\includegraphics[width=\columnwidth]{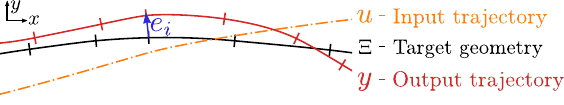} %
		\caption{A realization of Target, Input, and Output trajectories. The error is the distance between the output and the target. In the figure we plot $e_i$, the error for time step $i$. This is not to be confused with the ILC iteration number $k$. %
		}
		\label{fig:cartoon-simple}
	\end{center}
\end{figure}

Here we propose to design the policy $(x, \mathcal{T}, q)$ using SQP. We assume access to a model that is used to derive gradient and hessian information.
In subsequent sections we provide the details of the individual components in Fig.~\ref{fig:scheme}.
We initialize with a feasible input and take an SQP step after each experiment to evaluate the ILC policy~\eqref{eq:ideal_policy} using the approximate model of the system, measurements, and past inputs.
The objective of the approach is to minimize the output tracking error with respect to a target trajectory, illustrated in Fig.~\ref{fig:cartoon-simple}.

%% file: src/3-method.tex
\begin{figure}[t]
\vspace{0.2cm}
	\begin{center}
		\includegraphics[width=.7\columnwidth]{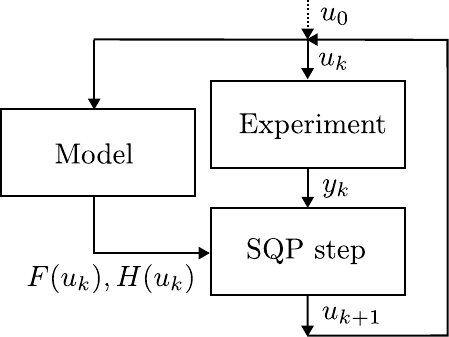} %
		\caption{
            Scheme of the proposed approach. We start with some initial input $u_0$.
            The output is obtained from an experiment.
            The model is used for the gradient and Hessian information in the SQP step, which is used for evaluating the next input.
        }
		\label{fig:scheme}
  
	\end{center}
\end{figure}

The process \eqref{eq:repetitive-process} is assumed to be nonlinear and unknown, as well its gradient and Hessian.
We assume it to be possible to evaluate \eqref{eq:repetitive-process}, and to have access to a model of the process either from first principles, experimental data or combinations of both, from which gradient and Hessian models can be derived. %
Our goal is to efficiently improve the quality of the output trajectory by leveraging the model information to reduce the number of experiments needed.

Driving the output of the repetitive process to the target geometry can be encoded as an optimization problem, where the cost function encodes the objective of tracking the target trajectory $\Target$ and the limits on input and output are framed as constraints.
The challenge is to incorporate process data into the optimization problem.
We can use tools from optimization theory to solve the nontrivial nonlinear constrained ILC problem.
In this work, we focus on adapting the SQP algorithm to compute the ILC updates.

We encode our control objective in the following optimization problem 
\begin{mini!}{z}{J(z)}{\label{eq:original-problem}}{}{\label{eq:original-problem-a}}%
	\addConstraint{h(z) \equiv p - f(u) }{= 0}{\label{eq:original-problem-b}}
	\addConstraint{g(z)}{\leq 0},
\end{mini!}%
where
$p$ is an optimization variable constrained by \eqref{eq:original-problem-b} to be equal to the noise-free system response $f(u)$,
$z = (u, p)$,
$\mathcal{U} \times \mathcal{Y} = \{z\,|\,g(z) \leq 0\}$, and
$J(z)$ is a function measuring the distance between the output and target trajectory (e.g., $J(z) = \|p - \Target\|$).

This is a nonlinear program, that we aim to solve using SQP. In SQP, we construct and solve a sequence of quadratic programs (QPs) that eventually converge to a solution of the original nonlinear problem. The Lagrangian associated with \eqref{eq:original-problem} is
\begin{equation}
	\mathcal{L}(z, \lambda, \sigma) = J(z) + \lambda^T h(z) + \sigma^T g(z),
\end{equation} 
and the standard quadratic subproblem is
\begin{mini!}{\Delta z}{\frac{1}{2} \Delta z\tp B \Delta z + \nabla J(z)\tp \Delta z }{\label{eq:qp-1}}{}
	\addConstraint{ \nabla h(z) \tp \Delta z + h(z)}{=0}
	\addConstraint{ \nabla g(z) \tp \Delta z + g(z)}{\leq 0},
\end{mini!} %
where $B \approx \nabla^2 \mathcal{L} (z, \lambda, \sigma)$ is an approximation for the Hessian of the Lagrangian.
We can use the primal-dual solution $(\Delta z^*,\lambda^*,\sigma^*)$  of the subproblem to construct the SQP-based ILC policy
\begin{equation}
	\mathcal{T}(z,\lambda,\sigma) = \begin{bmatrix}
	z + {\Delta z}^* \\
				\lambda^*\\
				\sigma^*
			\end{bmatrix}
\end{equation}

It is known that the iteration
\begin{equation}
	x_{k+1} = \mathcal{T}(x_k)
\end{equation}
where $x = (z,\lambda,\sigma)$ converges locally at a quadratic rate to minimizers of the original problem \eqref{eq:original-problem} that satisfy appropriate regularity conditions (e.g., the linear independence constraint qualification and strong second order sufficient conditions \cite{boggs1995sequential}).

We modify the SQP algorithm to design an ILC policy by incorporating data. 
	We assume that $f(u)$ is unknown but can be evaluated for any given $u_k$ by running an experiment leading to the output data $y_k = f(u_k) + w_k$  is corrupted by noise $w_k$.
Further, we assume to have access to approximations of the Jacobian and Hessian of the process derived from a system model
\begin{subequations}\label{eq:approximations}
\begin{align}
	F(u_k) & \approx \nabla f(u_k) \label{eq:gradient}\\ 
    H(u_k) & \approx \nabla^2 f(u_k) \label{eq:hessian} %
    \end{align}
\end{subequations}
We adapt \eqref{eq:qp-1} to deal with the fact that we do not have direct access to $f(u)$, replacing 
$h(z)$ with $p - y_k$, and
$\nabla h = [I~~-\nabla f]$ with $[I~~-F]$.
\if01
\begin{mini!}{\Delta z}{\frac{1}{2} \Delta z\tp \nabla^2 \mathcal{L}(z, \lambda, \sigma) \Delta z + \nabla J(z) \tp \Delta z}{\label{eq:qp-1}}{}
	\addConstraint{ [I~~-\nabla f] \tp \Delta z + p - y}{=0}
	\addConstraint{ \nabla g(z) \tp \Delta z + g(z)}{\leq 0}.
\end{mini!} %
\fi
\begin{mini!}{\!\Delta u, \Delta p}{\frac{1}{2}\! \dz\tp\! \begin{bmatrix} R & S \\ S\tp & Q\end{bmatrix} \!\dz\! \!+\!\! \begin{bmatrix} \nabla_u J \\ \nabla_p J \end{bmatrix}\tp \!\dz}{\label{eq:qp-2}}{\label{eq:qp-2a}}
	\addConstraint{ \Delta p }{= F(u_k) \Delta u + (y_k - p_k) \label{eq:qp-2b}}
    \addConstraint{ u_k + \Delta u}{\in \mathcal{U} \label{eq:qp-2c}}
    \addConstraint{ p_k + \Delta p}{\in \mathcal{Y} \label{eq:qp-2d}}
,\end{mini!} %
where
$R \approx \nabla^2_u \Lagrangian $, 
$S \approx \nabla^2_{y,u} \Lagrangian$, and 
$Q \approx \nabla^2_y \Lagrangian$.
The first term in \eqref{eq:qp-2b} imposes a linearized version of the dynamic constraint,
while the second term corrects the local estimate of the system output given the new measurement $y_k$.
Additionally, $R$ is constructed using the Hessian approximation $H(u_k)$.

Finally, we introduce a step size
$\eta_k$ as damping factor for the algorithm's iterates to encourage convergence; below we use a diminishing step size and provide simulation results on the effect of the decay rate. 
The resulting
\if01
iteration
\begin{equation}
    z_{k+1} = z_k + \eta_k \Delta z_k^*
\end{equation}
where $\Delta z^*$ is the solution to the modified data-dependant subproblem \eqref{eq:qp-2}
\else
OB-ILC policy then becomes
\begin{equation}
	\mathcal{T}(z,\lambda,\sigma, k) = \begin{bmatrix}
	   z_k + \eta_k \Delta z_k^* \\
	   \lambda^*\\
	   \sigma^*
	\end{bmatrix}
\end{equation}
where $(\Delta z^*, \lambda^*,\sigma^*)$ is the solution to the modified data-dependant subproblem \eqref{eq:qp-2} and $\Delta z^*=\left(\Delta u^*, \Delta p^* \right)$.
Note that constraint \eqref{eq:qp-2b} explicitely incorporates data from the real unknown system into our SQP algorithm to compensate for model mismatch and improve robustness.

\fi

The overall algorithm is outlined in Algorithm~\ref{alg:cap}.
The ILC loop is terminated when $\|\Delta z^*\|$ goes below a certain threshold, or when a maximum number of iterations is reached.
The SQP steps can be solved using a standard quadratic programming solver. In our implementation, we use OSQP~\cite{OSQP} via the Casadi~\cite{Casadi} interface for Python.
\begin{algorithm}
\caption{OB-ILC with SQP steps}\label{alg:cap}
\begin{algorithmic}[1]
\State $u \gets u_0$ \Comment{Initialization}
\Repeat
    \State $y_k \gets f(u_k) + w_k$ \Comment{Measurement}
    \State $\Delta z^*_k \gets$ Solution of \eqref{eq:qp-2}%
    \State $z_{k+1} \gets z_k + \eta_k \Delta z^*_k$ \Comment{Update}
    \State $k \gets k+1$
\Until{termination criteria is met}
\end{algorithmic}
\end{algorithm}

%% file: src/4-case-study.tex
In this section, we provide a detailed case study in precision tracking via a nonlinear high-fidelity simulator of a physical system.
\subsection{The system}

\begin{figure}[t]
\vspace{0.2cm}
	\begin{center}
		\includegraphics[width=\columnwidth]{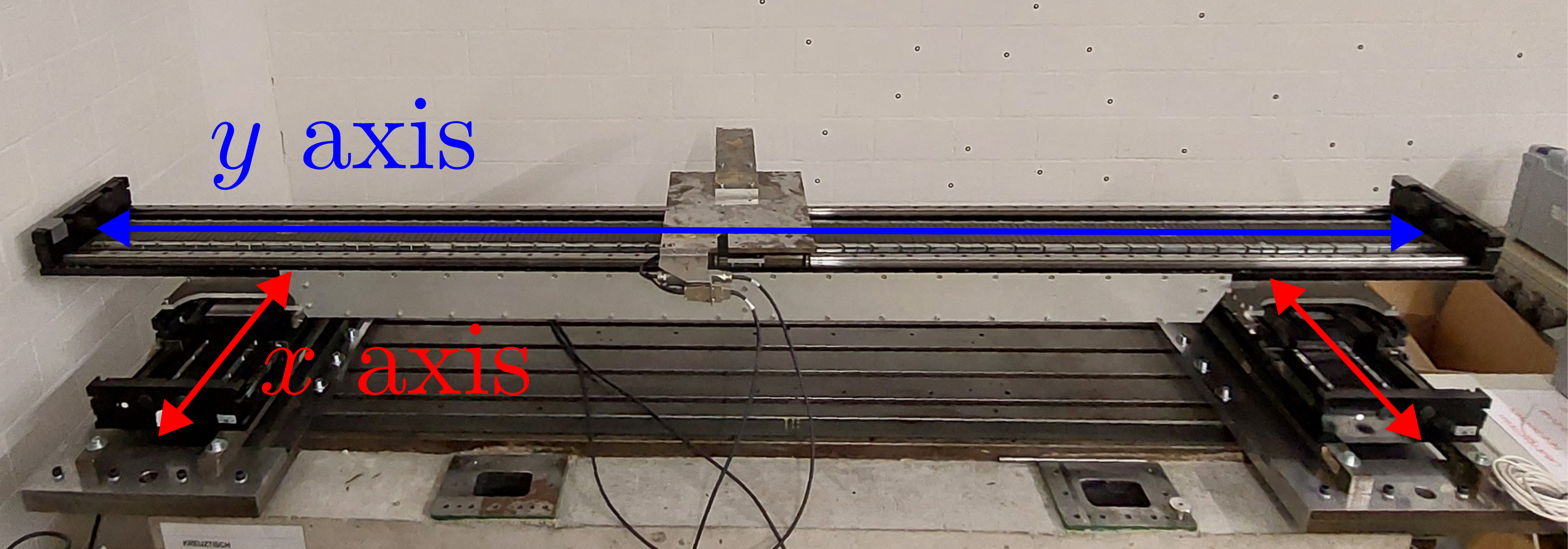} %
		\caption{2 axis precision motion system. Experimental data was collected from this system to build the models LM, NL1 and NL2.}
		\label{fig:andromeda}
	\end{center}
\end{figure}

As a case study we use a 2-axis high precision motion system depicted in Fig. \ref{fig:andromeda}. This system contains an internal closed-loop controller, it takes reference trajectories as inputs and produces tool-tip trajectories as outputs.
In this work, we use three different models to instantiate Algorithm~\ref{alg:cap}. All model the system response in discrete time, with a sample rate of $400\,\mathrm{Hz}$. For each case, we report $\sigma$, the standard deviation of the prediction error of the model, for input trajectories with acceleration up to $3\,\mathrm{m\,s^{-2}}$, when compared to experimental data. The three models are:
\begin{enumerate}
	\item[LM] A discrete-time linear model, in a state space lifted representation. $\sigma = 236.4\,\mathrm{\mu m}$.
	\item[NL1] A nonlinear ANN model, with an input layer capturing $200\,\mathrm{ms}$ of input history, and LeakyReLu activation functions. 
    $\sigma = 16.50\,\mathrm{\mu m}$.
	\item[NL2] A nonlinear ANN model, with an input layer capturing $500\,\mathrm{ms}$ of input history, and LeakyReLu activation functions. This model is, for the purposes of simulations, considered to be the ground truth. 
    $\sigma = 11.27\,\mathrm{\mu m}$.
\end{enumerate}
For a more detailed description of the system, model design and accuracy of the models see \cite{balula2022data}. 

In the following numerical results, we use either LM or NL1 to derive the gradient \eqref{eq:gradient} and Hessian \eqref{eq:hessian} information.
Since the structure of both models is kwown, one can take symbolic derivatives of the output in respect to the input to obtain the gradient and Hessian. %
The quality of the derivative information depends on the quality of the model derivatives, but it is not directly affected by the non-repeating disturbance, that is only used to determine the point of linearization. %
Due to the structure of both models, the Hessian evaluates to zero. In all simulations, the model NL2 is used in lieu of the true system, i.e., for the evaluations of $f(u)$ but not its gradients.

\subsection{Cost and constraints}
The optimization problem \eqref{eq:original-problem} used in the case study is
\begin{mini!}{u, p}{J(u, p) = \sum_{i = 0}^N \|p(i) - \Xi(i)\|^2_{Q_a}  + \|\partial^2 u(i) \|^2_{R_a}}{}{} %
\addConstraint{p-f(u)}{= 0}{}
\addConstraint{p(i)}{ \in \mathcal{W}}{,\quad i = 1, \dots, N }
\addConstraint{|\partial u(i)|}{ \leq \mathtt{v_{max} }}{,\quad i = 1, \dots, N - 2 }
\addConstraint{|\partial^2 u(i)|}{ \leq \mathtt{a_{max} }}{,\quad i = 1, \dots, N - 3 }
\addConstraint{|\partial^3 u(i)|}{ \leq \mathtt{j_{max} }}{,\quad i = 1, \dots, N - 4 }
,\end{mini!}
where $N$ is the number of points in the target trajectory $\Xi = \{\Xi_i\}_{i=1}^N \subseteq \Reals^2$,
$p = \{p_i\}_{i=1}^N \subseteq \Reals^2$ and $\mathcal{W} \subseteq \Reals^2$ is the workspace.
The first term of the cost function penalizes deviations of the output with respect to the target geometry, while the second term regularizes the input by penalizing the acceleration.
We use $Q_a = 10^{6} I$, $R_a=10^{-2} I$ to reflect the different order of magnitude of the input acceleration and output deviation. For the constraints we use 
$\mathtt{v_{max}} = 2$, $\mathtt{a_{max}} = 2$, $\mathtt{j_{max}} = 500$, derived from the physical limits of the machine. The number of points $N$ in the target trajectory (see Fig.~\ref{fig:xy-detail-linearmodel}) is $314$, which corresponds to $0.785\,\mathrm{s}$ time discretization between sample points given the sample rate of $400\,\mathrm{Hz}$. As the target geometry we use the outline of the letter 'r' from the ETH Zurich logo, shown in the inset of Fig. \ref{fig:xy-detail-linearmodel}.
The values for the velocity, acceleration, and jerk limits are derived from the physical limits of the machine.

%% file: src/5-results.tex
\subsection{Numerical Results}

\begin{figure}[t]
	\begin{center}
		\includegraphics[width=\columnwidth]{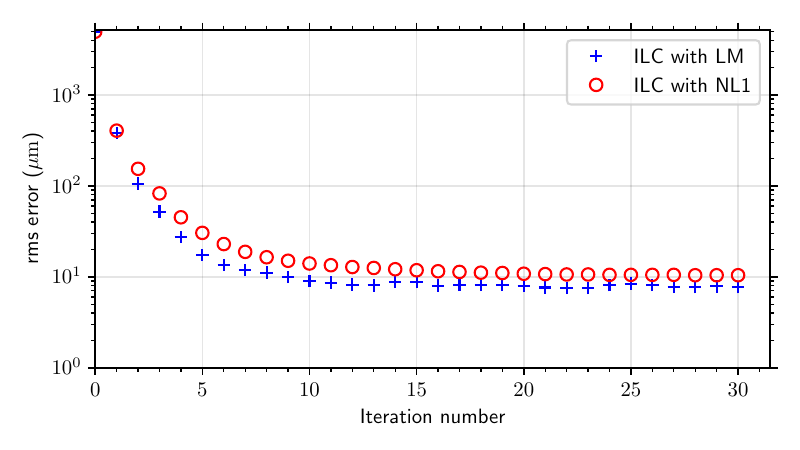} %
		\caption{Output deviation as a function of the iteration number for SQP steps using the derivative information from either the LM or NL1 models, evaluated with NL2 as ground truth.
  The initial condition for both series is the target geometry traced at constant velocity. Step size updated with $c=0.5$.}
		\label{fig:error-vs-iteration-init=xi}
	\end{center}
\end{figure}

\begin{figure}[t]
	\begin{center}
		\includegraphics[width=\columnwidth]{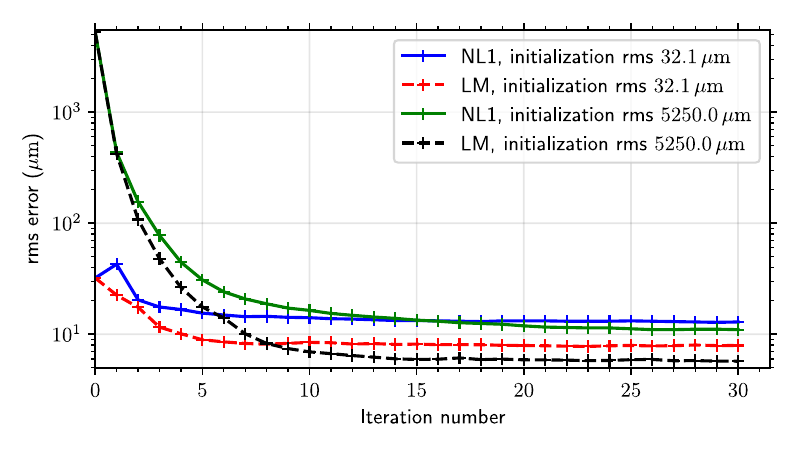} %
		\caption{Output deviation as a function of the iteration number for SQP steps using the derivative information from either the LM or NL1 models, evaluated with NL2 as ground truth.
                Two different initialization are provided for each case.
                The initialization starting with $rms=5250.0\,\mathrm{\mu m}$ is the solution of an optimal time problem with dynamics modeled with LM, where the velocity along the path is adjusted, for example negotiating intricate features slower and straight segments faster. The initialization starting with $rms=32.1\,\mathrm{\mu m}$ takes the optimal time solution as a starting point and solves a global optimization problem with dynamics modeled by NL1 and additional constraints on deviation, velocity, and acceleration. Further details can be found in \cite{balula2022data}. Step size $\eta_k = \eta_0 k^{-c}$ with $c=0.5$.}
		\label{fig:error-vs-iteration}
	\end{center}
\end{figure}

\begin{figure}[t]
\vspace{0.2cm}
	\begin{center}
		\includegraphics[width=\columnwidth]{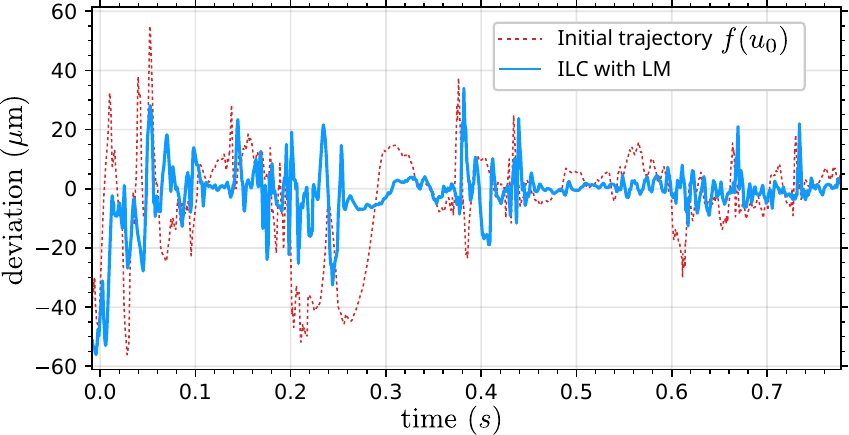} %
		\caption{Output error as a function of time, after $20$ ILC steps using the LM, and evaluated with NL2.
                }
		\label{fig:deviation-linearmodel}
	\end{center}
\end{figure}
\begin{figure}[t]
	\begin{center}
		\includegraphics[width=\columnwidth]{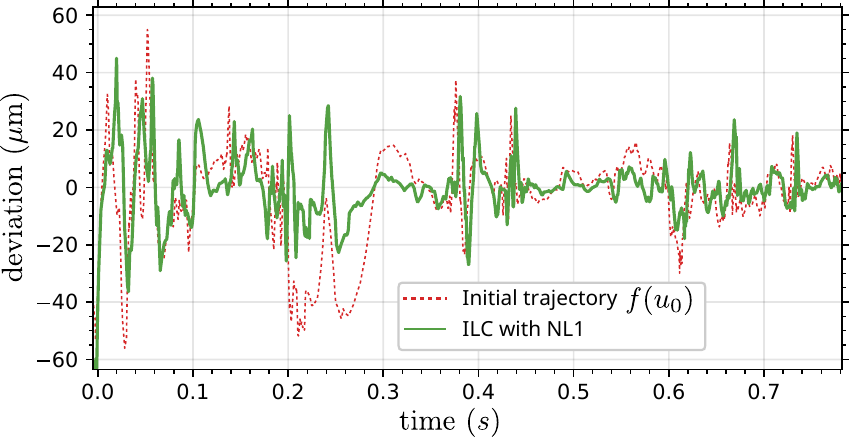} %
		\caption{Output error as a function of time, after $20$ ILC steps using the NL1, and evaluated with NL2.
                }
		\label{fig:deviation-nonlinearmodel}
	\end{center}
\end{figure}

\begin{figure}[t]
\vspace{0.2cm}
	\begin{center}
		\includegraphics[width=.7\columnwidth]{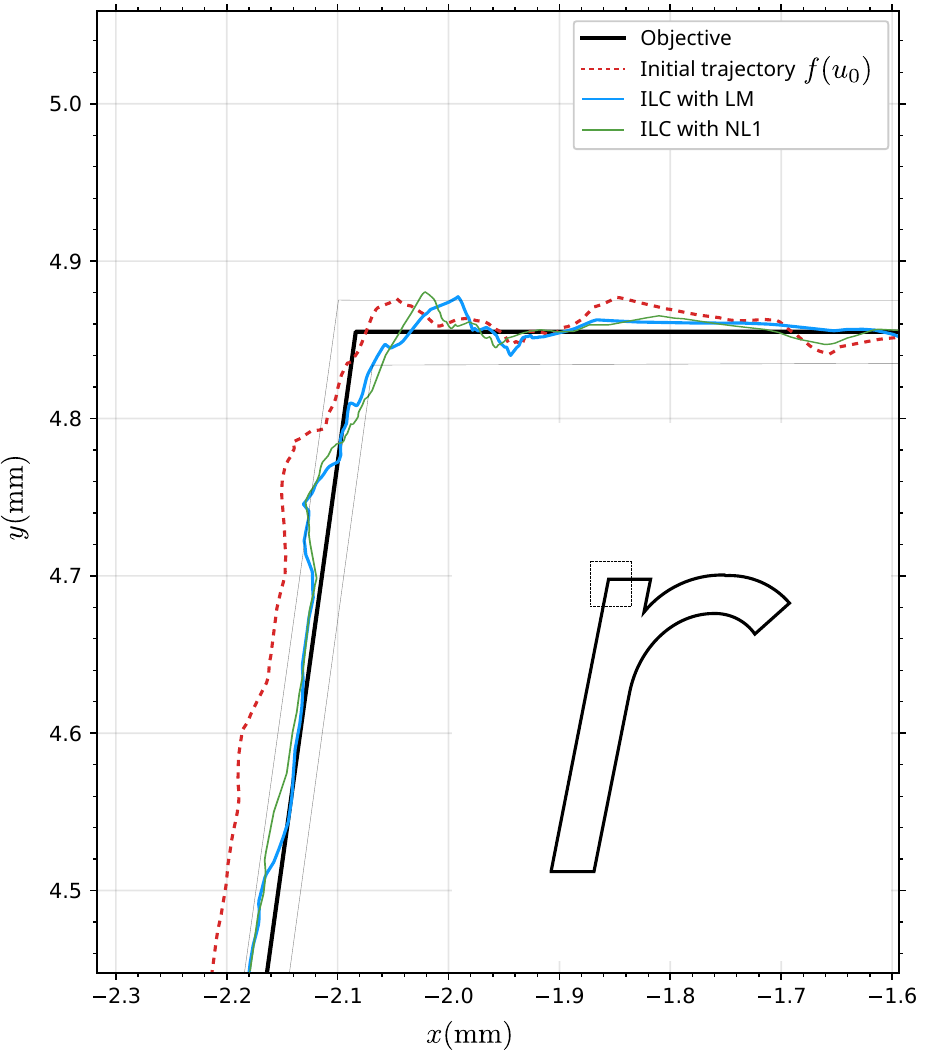} %
		\caption{Detail view of trajectories before and after $20$ ILC iterations. The figure shows a $20\,\mathrm{\mu m}$ band around the target geometry as a visual aid.
        }
		\label{fig:xy-detail-linearmodel}
	\end{center}
\end{figure}

\begin{figure}[t]
	\begin{center}
		\includegraphics[width=\columnwidth]{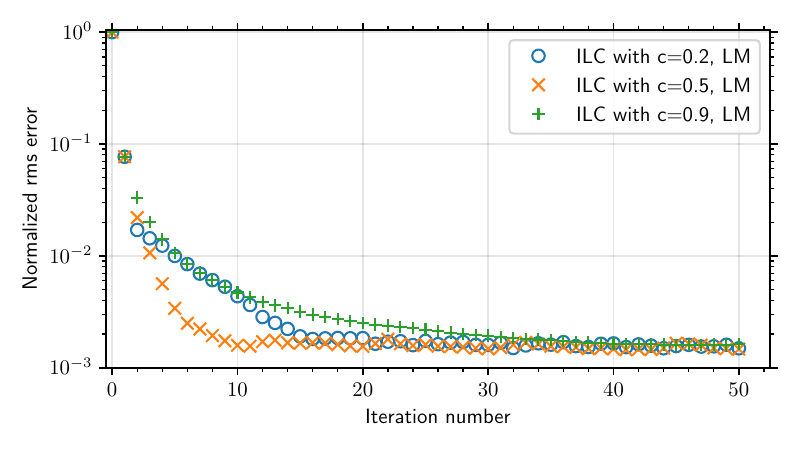} %
		\caption{Normalised error (rms) as a function of the iteration number for step size $\eta_k = \eta_0 k^{-c}$ with the LM, and an initialization with $\mathrm{rms} = 5250.0\,\mathrm{\mu m}$.}
		\label{fig:normalised-vs-iteration}
	\end{center}
\end{figure}

\if11
\begin{figure}[t]
\vspace{0.2cm}
	\begin{center}
		\includegraphics[width=\columnwidth]{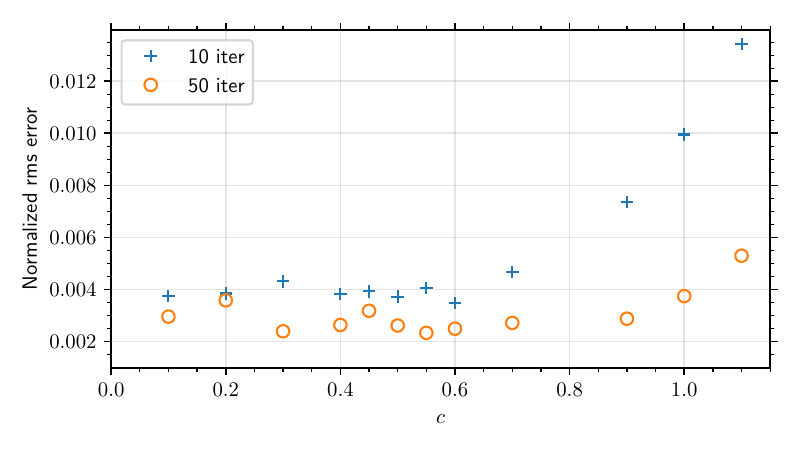} %
		\caption{Output error (rms) after $10$ and $50$ iterations for different step size rates $\eta_k = \eta_0 k^{-c}$ with the LM, and an initialization with $\mathrm{rms} = 5250.0\,\mathrm{\mu m}$.
                }
		\label{fig:rms-vs-c}
	\end{center}
\end{figure}
\fi

We present results using LM and NL1 for evaluating the approximations in \eqref{eq:approximations} and discuss the effect of 
the initial input $u_0$ and the decay rate of the form $\eta_k = \eta_0 k^{-c}$ on ILC performance.

\subsubsection{Effect of the model}
The results presented in Fig. \ref{fig:error-vs-iteration-init=xi} and Fig. \ref{fig:error-vs-iteration} show that for the iterations taken using gradient information from both the LM and the NL1, the error converges to a value of the same order of magnitude.
However, we see LM converge to a solution with lower deviation and at a faster rate compared to NL1.
This result is at first surprising given that the prediction error of the LM is one order of magnitude higher than NL1. We note however that NL1 is built with LeakyReLu activation functions, and thus its gradient is piecewise constant and discontinuous. Despite its higher prediction error, the structure of the LM is found to provide more accurate gradient information.
Different shapes and tunings of the cost function show mostly similar trends of the ILC loop using the linear and nonlinear models (data not shown).
We have observed that the system is only mildly nonlinear, and since the ILC step relies on measurements that do not depend on the models used, the fidelity of the approximations~\eqref{eq:approximations} is apparently not of critical importance. 
We hypothesize that a system with more pronounced nonlinearities would experience faster convergence with ILC steps relying on the nonlinear model for the approximations.
In Fig. \ref{fig:deviation-linearmodel} and Fig. \ref{fig:deviation-nonlinearmodel}, we show the output deviation as a function of time before and after $20$ steps of the proposed method. The deviations are computed between the output and the target trajectory depicted in Fig. \ref{fig:xy-detail-linearmodel}, where the output trajectories are plotted in $x-y$ coordinates.
Empirically, in previous studies, we found that the best rms error that can be obtained in this setup is close to the steady state values to which the ILC converges.
\subsubsection{Effect of initialization}
In Fig. \ref{fig:error-vs-iteration}, we show the results for two different initial conditions.
The results illustrate that since the underlying problem is nonlinear it is possible to be in a local minimum and not achieve a better solution, depending on the initial conditions.

\subsubsection{Effect of step size}

Next, we study the effect of step size on convergence behavior. We take $\eta_k = \eta_0 k^{-c}$ and compare the convergence.
We show the error trajectories for $c = \{0.2, 0.5, 0.9\}$ (Fig. \ref{fig:normalised-vs-iteration}).
In Fig. \ref{fig:rms-vs-c} we plot the error after $10$ and $50$ iterations for different values of $c$.
We observe that for $c$ between $0.3$ and $0.6$, we obtain fast decreases in the error without compromising the value at a steady state.

\if 01
\eb{
\begin{itemize}
    \item Comparison between different models starting from the same initial condition. 
    \item Compare to linear ILC and use the linear model for the Hessian etc. to show what we gain in simulation results
    \item Use the best-performing model and maybe do a plot on the effect of the initial condition to show the sensitivity to the initial condition, which should be part of the limitations of the method. Could make sense to compare against the linear ILC maybe? 
    \item (A side note that we could add only if we have time/space) If we have the 500ms NN as both model and system, we can show that the method (hopefully) almost converges to the optimization of the 500ms NN directly. This would be an additional motivation to use this framework for the optimization of such NN models instead of doing expensive optimization directly. 
\end{itemize}
}
\fi

%% file: src/6-conclusion.tex
We propose an algorithm for optimization-based ILC of nonlinear systems using sequential quadratic programming framework.
We formulate the ideal optimization problem and find approximate solutions using models of the true plant.
We illustrate the performance of the algorithm on a precision motion control simulation study using a high-fidelity simulator. The results show that our method works well under various parameter tunings and we are able to show significant improvement in the tracking error over baseline initial iterations. 

Future work will consider a formal analysis to provide robust convergence guarantees on the proposed method, as well as experiments on physical systems to verify the simulation study results.

%% file: main.bbl
\begin{thebibliography}{10}

\bibitem{liao2022robustness}
D.~Liao-McPherson, E.~C. Balta, A.~Rupenyan, and J.~Lygeros, ``On robustness in
  optimization-based constrained iterative learning control,'' {\em IEEE
  Control Systems Letters}, vol.~6, pp.~2846--2851, 2022.

\bibitem{barton2010norm}
K.~L. Barton and A.~G. Alleyne, ``A norm optimal approach to time-varying {ILC}
  with application to a multi-axis robotic testbed,'' {\em IEEE Transactions on
  Control Systems Technology}, vol.~19, no.~1, pp.~166--180, 2010.

\bibitem{son2015robust}
T.~D. Son, G.~Pipeleers, and J.~Swevers, ``Robust monotonic convergent
  iterative learning control,'' {\em IEEE Transactions on Automatic Control},
  vol.~61, no.~4, pp.~1063--1068, 2015.

\bibitem{tayebi2007unified}
A.~Tayebi and C.-J. Chien, ``A unified adaptive iterative learning control
  framework for uncertain nonlinear systems,'' {\em IEEE Transactions on
  Automatic Control}, vol.~52, no.~10, pp.~1907--1913, 2007.

\bibitem{adlakha2020optimization}
R.~Adlakha and M.~Zheng, ``An optimization-based iterative learning control
  design method for uav’s trajectory tracking,'' in {\em 2020 American
  Control Conference (ACC)}, pp.~1353--1359, IEEE, 2020.

\bibitem{mishra2010optimization}
S.~Mishra, U.~Topcu, and M.~Tomizuka, ``Optimization-based constrained
  {Iterative Learning Control},'' {\em IEEE Transactions on Control Systems
  Technology}, vol.~19, no.~6, pp.~1613--1621, 2010.

\bibitem{amann1996iterative}
N.~Amann, D.~H. Owens, and E.~Rogers, ``Iterative learning control for
  discrete-time systems with exponential rate of convergence,'' {\em IEE
  Proceedings-Control Theory and Applications}, vol.~143, no.~2, pp.~217--224,
  1996.

\bibitem{balta2021learning}
E.~C. Balta, K.~Barton, D.~M. Tilbury, A.~Rupenyan, and J.~Lygeros,
  ``Learning-based repetitive precision motion control with mismatch
  compensation,'' {\em arXiv preprint arXiv:2111.10246}, 2021.

\bibitem{gunnarsson2001design}
S.~Gunnarsson and M.~Norrl{\"o}f, ``On the design of ilc algorithms using
  optimization,'' {\em Automatica}, vol.~37, no.~12, pp.~2011--2016, 2001.

\bibitem{xu2011survey}
J.-X. Xu, ``A survey on iterative learning control for nonlinear systems,''
  {\em International Journal of Control}, vol.~84, no.~7, pp.~1275--1294, 2011.

\bibitem{yu2021neural}
Y.~Yu, C.~Zhang, Y.~Wang, and M.~Zhou, ``Neural-network-based iterative
  learning control for hysteresis in a magnetic shape memory alloy actuator,''
  {\em IEEE/ASME Transactions on Mechatronics}, vol.~27, no.~2, pp.~928--939,
  2021.

\bibitem{schollig2009optimization}
A.~Sch{\"o}llig and R.~D'Andrea, ``Optimization-based iterative learning
  control for trajectory tracking,'' in {\em 2009 European Control Conference
  (ECC)}, pp.~1505--1510, IEEE, 2009.

\bibitem{lu2017nonlinear}
J.~Lu, Z.~Cao, R.~Zhang, and F.~Gao, ``Nonlinear monotonically convergent
  {Iterative Learning Control} for batch processes,'' {\em IEEE Transactions on
  Industrial Electronics}, vol.~65, no.~7, pp.~5826--5836, 2017.

\bibitem{avrachenkov1998iterative}
K.~E. Avrachenkov, ``Iterative learning control based on quasi-newton
  methods,'' in {\em Proceedings of the 37th IEEE Conference on Decision and
  Control (Cat. No. 98CH36171)}, vol.~1, pp.~170--174, IEEE, 1998.

\bibitem{lin2006newton}
T.~Lin, D.~Owens, and J.~H{\"a}t{\"o}nen, ``Newton method based iterative
  learning control for discrete non-linear systems,'' {\em International
  Journal of Control}, vol.~79, no.~10, pp.~1263--1276, 2006.

\bibitem{volckaert2009model}
M.~Volckaert, A.~Van~Mulders, J.~Schoukens, M.~Diehl, and J.~Swevers, ``Model
  based nonlinear iterative learning control: A constrained gauss-newton
  approach,'' in {\em 2009 17th Mediterranean Conference on Control and
  Automation}, pp.~718--723, IEEE, 2009.

\bibitem{baumgartner2020zero}
K.~Baumg{\"a}rtner and M.~Diehl, ``Zero-order optimization-based iterative
  learning control,'' in {\em 2020 59th IEEE Conference on Decision and Control
  (CDC)}, pp.~3751--3757, IEEE, 2020.

\bibitem{boggs1995sequential}
P.~T. Boggs and J.~W. Tolle, ``Sequential quadratic programming,'' {\em Acta
  numerica}, vol.~4, pp.~1--51, 1995.

\bibitem{chen2021iterative}
Y.~Chen, B.~Chu, and C.~T. Freeman, ``Iterative learning control for
  path-following tasks with performance optimization,'' {\em IEEE Transactions
  on Control Systems Technology}, vol.~30, no.~1, pp.~234--246, 2021.

\bibitem{bolder2014rational}
J.~Bolder and T.~Oomen, ``Rational basis functions in iterative learning
  control—with experimental verification on a motion system,'' {\em IEEE
  Transactions on Control Systems Technology}, vol.~23, no.~2, pp.~722--729,
  2014.

\bibitem{bristow2006high}
D.~A. Bristow and A.-G. Alleyne, ``A high precision motion control system with
  application to microscale robotic deposition,'' {\em IEEE Transactions on
  Control Systems Technology}, vol.~14, no.~6, pp.~1008--1020, 2006.

\bibitem{OSQP}
B.~Stellato, G.~Banjac, P.~Goulart, A.~Bemporad, and S.~Boyd, ``{OSQP}: an
  operator splitting solver for quadratic programs,'' {\em Mathematical
  Programming Computation}, vol.~12, no.~4, pp.~637--672, 2020.

\bibitem{Casadi}
J.~A.~E. Andersson, J.~Gillis, G.~Horn, J.~B. Rawlings, and M.~Diehl,
  ``{CasADi} -- {A} software framework for nonlinear optimization and optimal
  control,'' {\em Mathematical Programming Computation}, vol.~11, no.~1,
  pp.~1--36, 2019.

\bibitem{balula2022data}
S.~Balula, D.~Liao-McPherson, A.~Rupenyan, and J.~Lygeros, ``Data-driven
  reference trajectory optimization for precision motion systems,'' {\em arXiv
  preprint arXiv:2205.15694}, 2022.

\end{thebibliography}
